\documentclass[12pt]{article}
\usepackage{hyperref}
\usepackage{graphicx}
\usepackage{amsmath}
\usepackage{bm}
\usepackage{amssymb}
\allowdisplaybreaks

\def\Li{\mathrm{Li}}
\newcommand{\z}[1]{\zeta_{#1}}
\newcommand{\sgnac}{[a^*]}
\newcommand{\sgna}{[a]}

\newcommand{\ZS}{\mathcal{Z}}
\newcommand{\wcZS}{\widetilde{\mathcal{Z}}}
\newcommand{\wZS}{\widetilde{Z}}
\newcommand{\wcS}{\widetilde{\mathcal{S}}}
\newcommand{\wS}{\widetilde{S}}
\newcommand{\tz}{\tilde{\zeta}}
\newcommand{\Imm}{\Im\mathrm{m}}
\newcommand{\Ree}{\Re\mathrm{e}}
\newcommand{\Ctl}{G}

\newcommand{\M}{N}

\begin{document}

\title{Analytic continuation of harmonic sums\\[2mm] with purely imaginary indices near the integer values}

\author{\sc V.N.~Velizhanin\\[5mm]
\it Theoretical Physics Division\\
\it NRC ``Kurchatov Institute''\\
\it Petersburg Nuclear Physics Institute\\
\it Orlova Roscha, Gatchina\\
\it 188300 St.~Petersburg, Russia\\[2mm]
\it velizh@thd.pnpi.spb.ru
}

\maketitle

\begin{abstract}
We present a simple algebraic method for the analytic continuation of harmonic sums with integer real or purely imaginary indices near negative and positive integers. We provide a MATHEMATICA code for exact expansion of harmonic sums in a small parameter near these integers. As an application, we consider the analytic continuation of the anomalous dimension of twist-1 operators in ABJM model, which contains the nested harmonic sums with purely imaginary indices. We found that in the BFKL-like limit the result has the same single-logarithmic behavior as in $\mathcal{N}=4$ SYM and QCD, however, we did not find a general expression for the ``BFKL Pomeron'' eigenvalue in this model. For the slope function, we found full agreement with the expansion of the known general result and give predictions for the first three perturbative terms in the expansion of the next-to-slope function. The proposed method of analytic continuation can also be used for other generalization of the nested harmonic sums.

\end{abstract}

\section{Introduction}\label{sec:intro}

The analytic continuation of the harmonic sums plays an important role in the study of a deep-inelastic process in the framework of the operator product expansion, where the nested harmonic sums enter into expressions for the coefficient functions and anomalous dimensions.

The harmonic sums defined recurrently as~\cite{Vermaseren:1998uu}
\begin{equation}
S_{a_1}(\M)
=\sum_{i_1=1}^\M \frac{({\rm sign}(a_1))^{i_1}}{i_1^{|a_1|}}\,,\qquad
S_{a_1,a_2,a_3,\ldots}(\M)
=\sum_{i_1=1}^\M \frac{({\rm sign}(a_1))^{i_1}}{i_1^{|a_1|}} S_{a_2,a_3,\ldots}(i_1)\,,
\label{HSR}
\end{equation}
for a positive integer  $\M$ can be extended to a complex argument in the same way that the simplest harmonic sum $S_1$ is related to the digamma function $\Psi$
\begin{eqnarray}
S_1(\M)&=&\sum_{i=1}^\M\frac{1}{i}=\sum_{i=1}^\infty\frac{1}{i}-\sum_{i=\M+1}^\infty\frac{1}{i}
=\sum_{i=1}^\infty\frac{1}{i}-\sum_{i=1}^\infty\frac{1}{i+\M}\nonumber\\
&=&\sum_{i=0}^\infty\frac{1}{i+1}-\sum_{i=0}^\infty\frac{1}{i+\M+1}
=\Psi(\M+1)-\Psi(1)\,,\label{S1}
\end{eqnarray}
where $\Psi(z)$ is defined as the logarithmic derivative of the gamma function
\begin{equation}
\Psi(z)=\frac{d}{dz}\ln\Big(\Gamma(z)\Big)=\frac{\Gamma'(z)}{\Gamma(z)}
\end{equation}
which has the single poles at $z=0,-1,-2,\ldots$. 
Such a procedure is usually called the analytic continuation and is described in details in Refs.~\cite{GonzalezArroyo:1979df,Kotikov:2005gr}. A general property of the analytically continued harmonic sum is the presence of poles in negative integer values of its argument. These poles are related to the evaluation equations for parton distributions, mostly to the Balitsky-Fadin-Kuraev-Lipatov (BFKL) equation~\cite{Lipatov:1976zz,Kuraev:1977fs,Balitsky:1978ic} and the double-logarithmic equation~\cite{Gorshkov:1966qd,Kirschner:1983di}. Knowledge of the analytic continuation of nested harmonic sums made it possible to find higher order corrections to the BFKL equation~\cite{Gromov:2015vua,Velizhanin:2015xsa,Velizhanin:2021bdh} and to the double-logarithmic equation~\cite{Velizhanin:2011pb,Velizhanin:2014dia,Velizhanin:2022seo}.

Analytic continuation of nested harmonic sums can be performed directly by following Refs.~\cite{GonzalezArroyo:1979df,Kotikov:2005gr}, as was done to compute the BFKL Pomeron eigenvalue in third order in Ref.~\cite{Gromov:2015vua} or as was used in Ref.~\cite{Albino:2009ci} to effective calculate the values of the nested harmonic sums for any complex argument (except negative integers).

Following another approach from Ref.~\cite{Blumlein:2009ta}, one can consider mappings the nested harmonic sums into their related Mellin transform functions and, using the relations between harmonic sums with a given weight, express them all in terms of several such basis functions that are analytic at $x=1$. Performing an asymptotic expansion of the basis functions for $\M\to\infty$ it is possible to obtain the numerical result for any complex value of $\M$.

In our previous work~\cite{Velizhanin:2020avm}, we proposed a more effective way to find the expansion of the nested harmonic sums near the negative integers by extractinf $\ln x$-terms in the inverse Mellin transform for the corresponding harmonic sums using \texttt{summer}~\cite{Vermaseren:1998uu} and \texttt{harmpol}~\cite{Remiddi:1999ew} packages for \texttt{FORM}~\cite{Vermaseren:2000nd,Kuipers:2012rf,Ruijl:2017dtg} along with database~\cite{Blumlein:2009cf}. The obtained database for analytic continuation of nested harmonic sums near negative and positive integers allowed us to generalised the double-logarithmic equation~\cite{Velizhanin:2022seo}. Knowing the pole expressions for the nested harmonic sums, one can use the dispersion representation~\cite{Velizhanin:2022ays} to obtain the value for any complex argument.

In this paper, we present a simple algebraic method that can be used for nested harmonic sums with real and purely imaginary indices. Such sums appear in the results for the anomalous dimension of composite operators in ABJM model~\cite{Lee:2017mhh,Lee:2019oml} and during the computations of the BFKL Pomeron eigenvalue in $\mathcal{N}=4$ SYM theory in the fourth order~\cite{Velizhanin:2021bdh}. The proposed method also works for generalisations of usual harmonic sums such as the so-called \texttt{SSum}~\cite{Moch:2001zr} or cyclotomic sums~\cite{Ablinger:2011te}.

\section{Analytic continuation}

We start our consideration with harmonic sum with one index $S_{a}$, which can be positive, negative or purely imaginary
\begin{equation}
S_a(\M)=\sum_{i=1}^{\M}\frac{\sgna^i}{i^{|a|}}\ \to\ 
\sum_{i=1}^{\infty}\frac{\sgna^i}{i^{|a|}}
-\sum_{i=1}^{\infty}\frac{\sgna^{i+\M}}{(i+\M)^{|a|}}\,,\qquad\quad 
\sgna\equiv \mathrm{sign}(a)= \frac{a}{|a|}
\label{Sa}
\end{equation}
where, for brevity, we used square brackets to denote the sign of index $a$.
The above expression is correct for positive integers $\M$. To make this expression correct for any complex $\M$, we must leave $\M$  only in the denominator, while in the numerator $\M$, which we will denote as $\M_0$ from now, is related to the initial positive integer value and gives in the case of usual alternating harmonic sums with the real indices~\cite{GonzalezArroyo:1979df,Kotikov:2005gr} a common plus/minus sign for even/odd $\M_0$ for the second term in Eq.~(\ref{Sa}).
The following expression
\begin{equation}
S_a(\M,\M_0)=
\sum_{i=1}^{\infty}\frac{\sgna^i}{i^{|a|}}
-\sgna^{\M_0}\sum_{i=1}^{\infty}\frac{\sgna^{i}}{(i+\M)^{|a|}}\label{SaAC}
\end{equation}
is correct now for any complex $\M$.

The last term near the negative integer $\M=-r+\omega$ is rewritten as follows:
\begin{eqnarray}
\sum_{i=1}^{\infty}\frac{\sgna^{i+N}}{\left(i-r+\omega\right)^{|a|}}
=
\sum_{i=1}^{r-1}\frac{\sgna^{i+N}}{\left(i-r+\omega\right)^{|a|}}
+\frac{\sgna^{r+N}}{\omega^{\,|a|}}
+\sum_{i=r+1}^{\infty}\frac{\sgna^{i+N}}{\left(i-r+\omega\right)^{|a|}}\,.\label{AC1}
\end{eqnarray}
The first term is related to harmonic sums (actually to Euler-Zagier sums, see below) due to the following property for integers $r$ and real or purely imaginary $a$:
\begin{equation}
\sum_{i=1}^{r-1}\frac{\sgna^{i}}{\left(i-r\right)^{|a|}}
=(-1)^{|a|}\,\sgna^{r}S_{a^*}(r-1)\label{ResummSa}
\end{equation}
where $a^*$ is complex conjugation.
When expanding Eq.~(\ref{AC1}) in $\omega$, the series will be sign-alternating, but for the first term, the sign will be compensated due to Eq.~(\ref{ResummSa}).

The general formula for the $\omega$-expansion near negative $\M=-r+\omega$ from even or odd $\M_0$ looks like:
\begin{equation}
S_{a}(\M,\M_0)=\zeta_a-\sgna^{r+\M_0}\left(\frac{1}{\omega^{|a|}}
+\sum_{k=0}\prod_{i=1}^k\big(|a^*|+k-1\big)\Big(S_{a^{*}+\sgnac k}(r-1)+(-1)^k\zeta_{a+[a]k}\Big)\,\frac{\omega^k}{k!}\right)\!.\label{SaNegExp}
\end{equation}
For the expansion near positive $\M$ we obtain directly from Eq.~(\ref{SaAC}) returning back to $S_a$
\begin{equation}
S_{a}(\M,\M_0)=\zeta_a-\frac{[a]^{\M_0}}{[a]^{\M}}\sum_{k=0}(-1)^k\prod_{i=1}^k\big(|a^*|+k-1\big)
\Big(-S_{a+[a]k}(\M)-\zeta_{a+[a]k}\Big)\,\frac{\omega^k}{k!}\,.\label{SaPosExp}
\end{equation}

For the harmonic sums with two indices, generalizing Eq.~(\ref{Sa}), we obtain
\begin{eqnarray}
S_{a_1,a_2}(\M,\M_0)&=&
\sum_{j=1}^{\infty}
\sum_{k=1}^{\infty}\Bigg[
\frac{[a_1]^{k+\M_0}}{(k+\M)^{|a_1|}}\frac{[a_2]^{j+k+\M_0}}{(j+k+\M)^{|a_2|}}
-\frac{[a_1]^{k+\M_0}}{(k+\M)^{|a_1|}}\frac{[a_2]^{j}}{j^{|a_2|}}\nonumber\\&&\qquad\qquad
-\frac{[a_1]^{k}}{k^{|a_1|}}\frac{[a_2]^{j+k}}{(j+k)^{|a_2|}}
+\frac{[a_1]^{k}}{k^{|a_1|}}\frac{[a_2]^{j}}{j^{|a_2|}}\Bigg]\qquad\label{UsualACSm21}
\\&&\hspace*{-30mm}=\
\sum_{j=1}^{\infty}
\sum_{k=1}^{\infty}
\frac{[a_1]^{k+\M_0}}{(k+\M)^{|a_1|}}\frac{[a_2]^{j+k+\M_0}}{(j+k+\M)^{|a_2|}}
-\zeta_{a_2}\sum_{k=1}^{\infty}\frac{[a_1]^{k+\M_0}}{(k+\M)^{|a_1|}}
-\zeta_{a_2,a_1}
+\zeta_{a_1} \zeta_{a_2}\label{UsualACSm21S}.
\end{eqnarray}
First term  
in the above equation, which we will call the generalised $\Psi$-function following Ref.~\cite{Kotikov:2005gr}, can be rewritten near negative integers $N=-r+\omega$~as:
\begin{eqnarray}
\frac{\Psi_{a_1,a_2}(r,\M_0)}{([a_1][a_2])^{\M_0}}&=&\sum_{i_1=1}^{\infty}\frac{[a_1]^{i_1}}{\left(i_1-r+\omega\right)^{|a_1|}}
\sum_{i_2=1}^{\infty}\frac{[a_2]^{i_1+i_2}}{\left(i_1+i_2-r+\omega\right)^{|a_1|}}
=\nonumber\\&
=&
\left(
\sum_{i_1=1}^{r-1}\frac{[a_1]^{i_1}}{\left(i_1-r+\omega\right)^{|a_1|}}
+\frac{[a_1]^{r}}{\omega^{|a_1|}}
+\sum_{i_1=r+1}^{\infty}\frac{[a_1]^{i_1}}{\left(i_1-r+\omega\right)^{|a_1|}}
\right)\times\nonumber\\&&\times
\left(
\sum_{i_2=1}^{r-1-i_1}\frac{[a_2]^{i_1+i_2}}{\left(i_1+i_2-r+\omega\right)^{|a_2|}}
+\frac{[a_2]^{r}}{\omega^{|a_2|}}
+\sum_{i_2=r-i_1+1}^{\infty}\frac{[a_2]^{i_1+i_2}}{\left(i_1+i_2-r+\omega\right)^{|a_2|}}
\right)\nonumber\\&
=&
\frac{[a_2]^{r}}{\omega^{|a_2|}}\sum_{i_1=1}^{r-1}\frac{[a_1]^{i_1}}{\left(i_1-r+\omega\right)^{|a_1|}}
+\frac{[a_1]^{r}}{\omega^{|a_1|}}
\sum_{i_2=1}^{\infty}\frac{[a_2]^{i_2+r}}{\left(i_2+\omega\right)^{|a_2|}}
\nonumber\\&&
+\sum_{i_1=1}^{r-1}\frac{[a_1]^{i_1}}{\left(i_1-r+\omega\right)^{|a_1|}}
\sum_{i_2=1}^{r-1-i_1}\frac{[a_2]^{i_1+i_2}}{\left(i_1+i_2-r+\omega\right)^{|a_2|}}
\nonumber\\&&
+\sum_{i_1=1}^{r-1}\frac{[a_1]^{i_1}}{\left(i_1-r+\omega\right)^{|a_1|}}
\sum_{i_2=1}^{\infty}\frac{[a_2]^{i_1+r}}{\left(i_2+\omega\right)^{|a_2|}}
+\sum_{i_1=1}^{\infty}\frac{[a_1]^{i_1+r}}{\left(i_1+\omega\right)^{|a_1|}}
\sum_{i_2=1}^{\infty}\frac{[a_2]^{i_1+i_2+r}}{\left(i_1+i_2+\omega\right)^{|a_2|}}
\nonumber\\&=& 
[a_1]^r[a_2]^r\bigg(
(-1)^{|a_1|}(-1)^{|a_2|}\wZS_{a_1^*,a_2^*}
+(-1)^{|a_1|}\wZS_{a_1^*}\tilde{\zeta}_{a_2}
\nonumber\\&&
+\tilde{\zeta}_{a_2,a_1}
+(-1)^{|a_1|}\wZS_{a_1^*}\frac{1}{\omega^{|a_2|}}
+\frac{1}{\omega^{|a_1|}}\tilde{\zeta}_{a_2}\bigg)\label{PsiaaNegExp}
\end{eqnarray}
where $\wZS_{\vec{a}}$ and $\tilde{\zeta}_{\vec{a}}$ denote the $\omega$-expansion, which generalize Eq.~(\ref{SaNegExp})
\begin{eqnarray}
\wZS_{a_1,a_2}&=&
{\ZS}_{a_1,a_2}
+\dot{\ZS}_{a_1,a_2}\omega
+\frac{1}{2}\ddot{\ZS}_{a_1,a_2}\omega^2
+\frac{1}{3!}\dddot{\ZS}_{a_1,a_2}\omega^3
+\frac{1}{4!}{\ZS}^{(4)}_{a_1,a_2}\omega^4+\cdots\\
\tilde{\zeta}_{a_1,a_2}&=&
{\zeta}_{a_1,a_2}
-\dot{\zeta}_{a_1,a_2}\omega
+\frac{1}{2}\ddot{\zeta}_{a_1,a_2}\omega^2
-\frac{1}{3!}\dddot{\zeta}_{a_1,a_2}\omega^3
+\frac{1}{4!}{\zeta}^{(4)}_{a_1,a_2}\omega^4+\cdots
\label{SaaNegExp}
\end{eqnarray}
that is, a Tailor expansion with a formal ``differentiation'' with respect to indices, which we define as
\begin{equation}
\dot{F}_{a_1,a_2,\ldots,a_n}=
|a_1|F_{a_1+[a_1],a_2,\ldots,a_n}
+|a_2|F_{a_1,a_2+[a_2],\ldots,a_n}
+\cdots
+|a_n|F_{a_1,a_2,\ldots,a_n+[a_n]}
\end{equation}
Note, that for the expansion of the Euler-Zagier sums $\ZS_{\vec{a}}$ the series is not sign-alternating, as for $\zeta_{\vec{a}}$, due to Eq.~(\ref{ResummSa}). In Eq.~(\ref{UsualACSm21S}) we used the following properties for the term with finite summations
\begin{equation}
\sum_{i_1=1}^{r-1}\frac{[a_1]^{i_1}}{\left(i_1-r+\omega\right)^{|a_1|}}
\sum_{i_2=1}^{r-1-i_1}\frac{[a_2]^{i_1+i_2}}{\left(i_1+i_2-r+\omega\right)^{|a_2|}}=
[a_1]^r[a_2]^r(-1)^{|a_1|}(-1)^{|a_2|}\ZS_{a_1^*,a_2^*}
\end{equation}

For the expansion near positive $\M$, we obtain, by generalizing Eq.~(\ref{SaPosExp}), for the first term in Eq.~(\ref{UsualACSm21S})
\begin{eqnarray}
S_{a_1,a_2}(\M,\M_0)&=&
{\zeta}_{a_1}{\zeta}_{a_2}
-{\zeta}_{a_2,a_1}
-\frac{[a_1]^{\M_0}}{(a_1)^{\M}}\left(-\widetilde{S}_{a_1}
-\tilde{\zeta}_{a_1}
\right){\zeta}_{a_2}\nonumber\\&&
+\frac{([a_1][a_2])^{\M_0}}{([a_1][a_2])^{\M}}\left(\widetilde{S}_{a_1,a_2}
-\widetilde{S}_{a_1}\tilde{\zeta}_{a_2}
+\tilde{\zeta}_{a_2,a_1}\right)\,,
\label{PsiaaPosExp}
\end{eqnarray}
where the last term without common prefactor is the analytic continuation for $\Psi_{\vec{a}}(\M,\M_0)$ near positive integers $\M$. It can be seen that there is a minus sign in the brackets before the harmonic sum with one index or, in general, with an odd number of indices and this is a general properties due to transformation from the usual harmonic sum to the combination of $\Psi_{\vec{a}}$ and $\zeta_{\vec{c}}$ and vice versa.

The multiple zeta values $\zeta_{\vec{a}}$ can be reduced to several basis MZV's using their relations with harmonic polylogarithms~\cite{Remiddi:1999ew} or with multiple polylogarithms~\cite{Goncharov:MPL}. In the first case, we can use the existing database for the harmonic polylogarithms at unity $H_{\vec{c}}(1)$ ~\cite{Blumlein:2009cf} and the following relation between the vectors $\vec{a}$ and $\vec{c}$ in $\zeta_{\vec{a}}$ and $H_{\vec{c}}(1)$
\begin{equation}
\{a_1,a_2,a_3,\ldots,a_n\}\to
\Big\{a_1\prod_{i=1}^1[a_i],a_2\prod_{i=1}^2[a_i],a_3\prod_{i=1}^3[a_i],
\ldots,a_n\prod_{i=1}^n[a_i]\Big\}
\end{equation}
and multiplying $H_{\vec{c}}(1)$ by the additional sign-factor
\begin{equation}
\zeta_{\vec{a}}=\Li_{|\vec{a}|}([\vec{a}])=\prod_{i=1}[a_i]\, H_{\vec{c}}(1)
\end{equation}
However, the database ~\cite{Blumlein:2009cf} only contains results for real indices (positive and negative integers). In the case of purely imaginary indices, it was necessary to obtain such a database, which was done in \cite{Velizhanin:I}, where we computed the numerical values of multiple polylogarithms with high accuracy using the GiNaC~\cite{Bauer:2000cp} implementation of their numerical evaluation~\cite{Vollinga:2004sn} 
and found the relationship between multiple polylogarithms with several basis ones using PSLQ method~\cite{Ferguson1999AnalysisOP}\footnote{We used \texttt{MATHEMATICA} function \texttt{FindIntegerNullVector} for lowest weights and code \cite{Zimmermann:CPSLQ} for higher weights up to weight 7. In the forthcoming paper~\cite{Velizhanin:I} we will extend this database up to weight 8.}.

One can numerically check the agreement between the first term in Eq.~(\ref{UsualACSm21S}) and Eqs.~({\ref{PsiaaNegExp}) and~({\ref{PsiaaPosExp}) using the program described in the next Section.

In general, the analytic continuation can be subdivided into two steps. First of all, we should move from the usual nested harmonic sums $S_{\vec{a}}$ to the combination of the $\Psi_{\vec{a}}$ and $\zeta_{\vec{a}}$. Then we perform the expansion of $\Psi_{\vec{a}}$
as in Eq.~(\ref{PsiaaNegExp}) or in Eq.~(\ref{PsiaaPosExp}).

In the case of three indices, the generalization of Eq.~(\ref{SaaNegExp}) for $\M=-r+\omega$ looks like
\begin{eqnarray}
\widetilde{\Psi}_{a_1,a_2,a_3}(r,\M_0)&=&
\frac{\Psi_{a_1,a_2,a_3}(r,\M_0)}{\prod_{i=1}^3[a_i]^r}=
\wcZS_{a_1,a_2,a_3}
+\wcZS_{a_1,a_2}\tz_{a_3}\nonumber\\&&
+\wcZS_{a_1}\tz_{a_3,a_2}
+\tz_{a_3,a_2,a_1}
+\wcZS_{a_1,a_2}\frac{[a_3]^r}{\omega^{|a_3|}}
+\wcZS_{a_1}\frac{[a_2]^r}{\omega^{|a_2|}}\tz_{a_3}
+\frac{[a_1]^r}{\omega^{|a_1|}}\tz_{a_3,a_2}
\end{eqnarray}
with replacement
\begin{equation}
\wcZS_{\vec{a}}=(-1)^{w_{\vec{a}}}\wZS_{\vec{a}}\,,
\end{equation}
where $w_{\vec{a}}=\sum_{\vec{a}}|a_i|$ is the weight of Euler-Zagier sum $\ZS_{\vec{a}}$.
while the generalization of Eq.~(\ref{PsiaaPosExp}) for positive $\M$ is the following
\begin{eqnarray}
\Psi_{a_1,a_2,a_3}(\M,\M_0)&=&\frac{\prod_{i=1}^3[a_i]^{\M_0}}{\prod_{i=1}^3[a_i]^{\M}}
\bigg(-\wS_{a_1,a_2,a_3}+\wS_{a_1,a_2}\tz_{a_3}-\wS_{a_1}\tz_{a_3,a_2}+\tz_{a_3,a_2,a_1}\bigg)
\end{eqnarray}

In general case, the corresponding expansions can be written as
\begin{eqnarray}
\widetilde{\Psi}_{a_1,a_2,a_3,\ldots,a_n}
&=&
\wZS_{a_1,a_2,a_3,\ldots,a_n}
+\wZS_{a_1}\tz_{a_n,\ldots,a_3,a_2}
+\wZS_{a_1,a_2}\tz_{a_n,\ldots,a_3}
+\cdots
+\tz_{a_n,\ldots,a_3,a_2,a_1}\nonumber\\&&
+\frac{1}{\omega^{|a_1|}}\tz_{a_{n},\ldots,a_3,a_2}
+\wZS_{a_1}\frac{1}{\omega^{|a_2|}}\tz_{a_{n},\ldots,a_3}
+\cdots
+\wZS_{a_1,a_2,a_3,\ldots,a_{n-2}}\frac{1}{\omega^{|a_{n-1}|}}\tz_{a_{n}}\nonumber\\&&
+\wZS_{a_1,a_2,a_3,\ldots,a_{n-1}}\frac{1}{\omega^{|a_n|}}
\end{eqnarray}
and
\begin{eqnarray}
\Psi_{a_1,a_2,a_3,\ldots,a_n}(\M,\M_0)
=\frac{\prod_{i=1}^n[a_i]^{\M_0}}{\prod_{i=1}^n[a_i]^{\M}}
\bigg(&&\wcS_{a_1,a_2,a_3,\ldots,a_n}
+\wcS_{a_1}\tz_{a_n,\ldots,a_3,a_2}\nonumber\\&&
+\wcS_{a_1,a_2}\tz_{a_n,\ldots,a_3}
+\cdots
+\tz_{a_\ell,\ldots,a_3,a_2,a_1}\bigg)
\end{eqnarray}
with replacement
\begin{equation}
\wcS_{\vec{a}}=(-1)^{\ell_{\vec{a}}}\wS_{\vec{a}}\,,
\end{equation}
where $\ell_{\vec{a}}=\ell_{a_1,a_2,\ldots,a_n}=n$ is the number of indices in nested harmonic sum $S_{\vec{a}}$ or  $\Psi_{\vec{a}}$.
So, we have the rather simple and easy-to-program algebraic method for the analytic continuation of the nested harmonic sums, and in the next Section we provide the code that implements the proposed method.

\section{Implementation in \texttt{MATHEMATICA} }\label{MATH}

We implemented the above described method in \texttt{MATHEMATICA}. All functions discussed in this Section are collected in \texttt{ACHSI.m} file\footnote{This code is available as ancillary files of \texttt{arXiv}-version of this paper and on GitHub \url{https://github.com/vitvel/ACHSI}.} and it must be preloaded in \texttt{MATHEMATICA} session as usual
\begin{verbatim}
<<"ACHSI.m";
\end{verbatim}
adding the full path, if necessary. 

In the first step, we convert the nested harmonic sum $S_{\vec{a}}$ into an expression containing $\Psi_{\vec{a}}$ and $\zeta_{\vec{a}}$. This can be done using the function \texttt{SToPsi}:
\begin{eqnarray}
\mathtt{SToPsi[}S_{-1,2i,-3}\mathtt{]}&=&
\zeta_{-3, 2i, -1}
- \Psi_{-1, 2i, -3}
+ \zeta_{-3} \Psi_{-1, 2i}
- \zeta_{-3} \zeta_{2i, -1}\nonumber\\&&
+ \zeta_{-1} \zeta_{-3} \zeta_{2i} 
-\Psi_{-1} \zeta_{-3} \zeta_{2 i}
+ \Psi_{-1} \zeta_{-3, 2i}
- \zeta_{-1} \zeta_{-3, 2i}\,.
\end{eqnarray}
The analytic continuation of $\Psi_{\vec{a}}$ can be performed with the function \texttt{PsiAC} for the $\omega$-expansion near negative and positive integers, which has four arguments: $\Psi_{\vec{a}}$, final value $\M$, initial value $\M_0$ and the order of $\omega$-expansion. 
Since the analytic continuation contains $\zeta_{\vec{a}}$ one should load preliminary the database for the relations between $\zeta_{\vec{a}}$ or their associated multiple polylogarithms 
\begin{equation}
\zeta_{\vec{a}}=\Li_{|\vec{a}|}([\vec{a}])\,,
\end{equation}
and we use the standard definition of the multiple polylogarithms
\begin{equation}
\Li_{m_1,\ldots,m_k}(x_1,\ldots,x_k)=\sum_{i_1>i_2>\ldots>i_k>0}
\frac{x_1^{i_1}}{i_1^{m_1}}
\cdots
\frac{x_k^{i_k}}{i_k^{m_k}}\,.
\end{equation}
We have such database for all $\zeta_{\vec{a}}$ with positive, negative and purely imaginary indices (or for multiple polylogarithms with positive, negative and purely imaginary arguments) up to weight~6 and with positive, negative and only positive (or only negative related with positive by complex conjugation) 
purely imaginary indices for weight~7\footnote{The results for the NNNLLA BFKL Pomeron eigenvalue contains harmonic sums only with one purely imaginary index.}. The precomputed database should be loaded inside a \texttt{MATHEMATICA} session prior to using the \texttt{PsiAC} function with the following code up to weight~6
\begin{verbatim}
DLiSubsRe = Get["LiSubsRew1w6.m"]//Dispatch;
DLiSubsIm = Get["LiSubsImw1w6.m"]//Dispatch;
\end{verbatim}
or up to weight 7
\begin{verbatim}
DLiSubsRe = Join[Get["LiSubsRew1w6.m"],Get["LiSubsRew7.m"]]//Dispatch;
DLiSubsIm = Join[Get["LiSubsImw1w6.m"],Get["LiSubsImw7.m"]]//Dispatch;
\end{verbatim}
adding the full path, if necessary.

However, $\zeta_{\vec{a}}$ starting with index equal to 1, $\zeta_{1,\vec{a}}$, diverge and require separate consideration. For this purpose, we use the so-called stuffle relation for multiple polylogarithms, which can be written in the case of two indices as
\begin{equation}
\Li_{i_1}(x_1)\Li_{i_2}(x_2)=
 \Li_{i_1,i_2}(x_1,x_2)
+\Li_{i_2,i_1}(x_2,x_1)
+\Li_{i_1+i_2}(x_1 x_2),
\end{equation}
in the case of three indices as
\begin{eqnarray}
\Li_{i_1}(x_1)\Li_{i_2,i_3}(x_2,x_3)&=&
 \Li_{i_1,i_2,i_3}(x_1,x_2,x_3)
+\Li_{i_2,i_1,i_3}(x_2,x_1,x_3)
+\Li_{i_2,i_3,i_1}(x_2,x_3,x_1)\nonumber\\&&
+\Li_{i_1+i_2,i_3}(x_1 x_2,x_3)
+\Li_{i_2,i_1+i_3}(x_2,x_1 x_3)
\end{eqnarray}
and can be easily generalized to the case of any number of indices. Such relations allow us to extract $\Li_1(1)$, which can be performed with the function \texttt{Li1Extract}
\begin{verbatim}
Li1Extract[Li[1, 1, 2 I]] =
   1/2*Li1^2*Li[2 I] - Li1*Li[3 I]  - Li1*Li[2 I, 1] + 1/2*Li[4 I] + 
   1/2*Li[2 I, 2] + Li[3 I, 1] - 1/2*Li[2, 2 I] + Li[2 I, 1, 1]
\end{verbatim}
and this function is contained within \texttt{PsiAC} function.

After loading the precomputed database, we can compute the $\omega$-expansion of the $\Psi_{\vec{a}}$ with the function \texttt{PsiAC} as
\begin{verbatim}
PsiAC[S[2,-1,I],-3,1,0]
\end{verbatim}
which produces the following general expression for the analytic continuation of $\Psi_{2,-1,i}$ from odd positive $(\M_0\pmod{4})=1$ to odd negative $(\M\pmod{4})=1$ up to $\omega^0$
\begin{eqnarray}
&&
\frac{1}{\omega^2}\bigg(
\frac{\pi^2}{32}
+ \frac{\ln\!2^2}{8}
+ i\Big( \Ctl 
- \frac{3}{8} \pi \ln\!2\Big)
\bigg)
+ \frac{1}{\omega}\Big(
  S_{-3}
- S_{2, -1}
- \frac{1}{96} \pi^2\ln\!2\nonumber\\&&
- \frac{27}{64} \z3
- \frac{1}{2}  S_{2}\ln\!2 
+ i\Big(
  \frac{3}{4} \Ctl \ln\!2
- \frac{13}{384}\pi^3
- \frac{3}{2} \Imm(\Li_{i, 2})
+ \frac{1}{4} \pi S_2 
\Big) \nonumber\\&&
- S_{ 2, 2 i}
- 2 S_{3, -1}
+ S_{ 2, -1, -i}
+ 3 S_{-4}
+ S_{4 i}
- S_{-3, -i}
- S_{2, -2}
+ \frac{1}{2}\Ctl^2\nonumber\\&&
+ \frac{3}{16}\Ctl  \pi\ln\!2
- \frac{1}{384}\pi^2\ln\!2^2  
+ \frac{253}{46080}\pi^4
- \frac{5}{4} \z3 \ln\!2
- \frac{3}{8} \pi \Imm(\Li_{i, 2}) \nonumber\\&&
+ \frac{45}{32} \Li_{-3, 1}
- \frac{1}{2}S_{-3}\ln\!2 
+ \frac{1}{8} S_{2}\ln\!2^2
+ \frac{5}{96} S_2 \pi^2 
- S_3\ln\!2  
+ \frac{1}{2} S_{2, -1} \ln\!2\nonumber\\&&
+i\Big(
\frac{37}{144} \Ctl \pi^2 
- \frac{1}{6} \Ctl \ln\!2^2 
- \frac{29}{576} \pi^3 \ln\!2 
- \frac{47}{384} \pi \z3
- \frac{4}{3} \Imm(\Li_{4 i})\ln\!2\nonumber\\&&
+ \frac{1}{3}  \Imm(\Li_{i, 2})
+ \frac{5}{3} \Imm(\Li_{2, 1, i})
+ \frac{1}{4}  S_{-3}\pi	
- \frac{3}{8}  S_2 \pi \ln\!2
+ \frac{1}{2} S_3 \pi 
- \frac{1}{4} S_{2, -1}\pi\, ,
\end{eqnarray}
where we converted the Euler-Zagier sums to usual harmonic sums by calling the function \texttt{ZSToHS} inside \texttt{PsiAC} and in the above expression $S_{\vec{a}}=S_{\vec{a}}(|\M|-1)$, while $\Ctl$ is the Catalan constant. Note, that these are general properties of analytic continuation for harmonic sums with real and purely imaginary indices, that the same general results will be obtained for $(\M_0\pmod{4})$, for $(\M\pmod{4})$ and, moreover, for $((\M_0+\M)\pmod{4})$. To obtain the analytic continuation into $\M=-3$, we need to put $S_{\vec{a}}=S_{\vec{a}}(|-3|-1)=S_{\vec{a}}(2)$ and use substitution for nested harmonic sums defined in \texttt{ACHSI.m} file, e.g.
\begin{verbatim}
S[2,-1,I] /. S[a__] :> HS[a, 2] = -1/16 - 9/8 I
\end{verbatim}

The analytic continuation of the nested harmonic sum can be performed with the function \texttt{HSAC}, which has the same arguments, as \texttt{PsiAC} function, and applying it to $S_{2,2i}$
\begin{verbatim}
HSAC[S[2, 2 I], -3, 1, 1]
\end{verbatim}
we get the following expression
\begin{eqnarray}&&
\frac{S_2}{\omega^2}
+ \frac{3/16 \z3 + 2 S_3-i \pi^3/16}{\omega}
+ 3 S_4 
- S_{-4i}
+ S_{2, -2 i}
-\frac{7}{3840} \pi^4
+ 3 i \Imm(\Li_{4i})\nonumber\\&&
+ \omega \Big[
4 S_5 
+ 2 S_{2, -3 i}
+ 2 S_{3, -2 i}
- 4 S_{-5 i}
+ \frac{3}{16} \z3 S_2
- 2 \z5 
+\frac{1}{16 }\Ctl \pi^3
+ \frac{35}{768 } \pi^2 \z3\label{S2m1IExp}\\&&\qquad
+i\Big(
2 \Ctl \z3 
- \frac{1}{16}\pi^3 S_2
- \frac{7 }{512 }\pi^5
+ \ln\!2\; \Imm(\Li_{4i}) 
+ 4 \Imm(\Li_{4i,1}) 
- 2 \Imm(\Li_{4,i}) 
\Big)
\Big]\nonumber
\end{eqnarray}
where again $S_{\vec{a}}=S_{\vec{a}}(|\M|-1)$.

When evaluating an expression involving several nested harmonic sums, as in real anomalous dimension, it is better to use substitution
\begin{verbatim}
S[2, 2 I] /. S[a__] :> HSAC[S[a], -3, 1, 1]
\end{verbatim}
with the same output as in Eq.~(\ref{S2m1IExp}).

For the numerical evaluation of the obtained results we provide the numerical values for the basis of MZV's with an accuracy of 50 digits in the form of substitution rules \texttt{LiNSubs} and it can be used as
\begin{verbatim}
HSAC[S[2,2I],-3,1,1]/.LiNSubs/.S[a__]:>HS[a,Abs[-3]-1]//N[#,5]&
\end{verbatim}
with the output
\begin{equation}
\frac{1.25000 + 0. 10^{-6} i}{\omega^2}
+ \frac{2.4754 - 1.9379 i}{\omega}
+(3.0099 + 2.7168 i) 
+ (4.6486 - 4.6926 i)\, \omega\,.
\end{equation}
In particular, one can numerically check the agreement between the first term in Eq.~(\ref{UsualACSm21S}) and Eqs.~({\ref{PsiaaNegExp}) and~({\ref{PsiaaPosExp}).

\section{Application}\label{Sec:Application}

In this Section, we provide several examples of using \texttt{HSAC} function to study the relevant quantities in various models. The main motivation for studying analytic continuation with the real and purely imaginary indices was the computation of the BFKL eigenvalue in the forth order~\cite{Velizhanin:2021bdh}, using the Quantum Specral Curve (QSC) method~\cite{Gromov:2013pga,Gromov:2014caa,Gromov:2015vua}. During the computations we found an inconsistency when using the usual nested harmonic sums. We extended our consideration to include nested harmonic sums with purely imaginary indices, since similar objects appeared in the results for the anomalous dimensions of composite operators in ABJM model~\cite{Aharony:2008ug} calculated by the same QSC method in Refs.~\cite{Lee:2017mhh,Lee:2019oml}. We found that considering nested harmonic sums (more precisely, functions $\Psi_{\vec{a}}$) with the last purely imaginary index will considerably simplify the final result and resolve the initial inconsistency.

Another example where nested harmonic sums with real and purely imaginary indices appear is the above-mentioned anomalous dimension of composite operators in ABJM model.
The available results up to third order are expressed in terms of the following nested harmonic sums~\cite{Lee:2017mhh,Lee:2019oml}
\begin{equation}
H_{a}(\M)=\sum_{k=1}^\M\frac{\Ree([a]^k)}{k^{|a|}}\,,\qquad
H_{a_1,a_2,\ldots,a_n}(\M)=\sum_{k=1}^\M\frac{\Ree([a_1]^k)}{k^{|a_1|}}H_{a_2,\ldots,a_n}(k)\,.
\end{equation}
with negative and positive real integer and only positive purely imaginary integer indices. To return to the usual definition of nested harmonic sums, we must also consider the negative purely imaginary indices and replace each $i$ in $H_{\vec{a}}$ by the sum of $S_{\vec{a}}$ with $\pm i$ like
\begin{eqnarray}
H_i=\frac{1}{2}\Big(S_i+S_{-i}\Big)
\end{eqnarray}
Inside \texttt{MATHEMATICA} such a transformation can be done with the following code
\begin{verbatim}
 H[I,2I,3I]/.H[a__]:>HS[a]//.HS[a___,b_/;Sign[b]==I, c___]:>
    (HS[a,Abs[b]II,c]+HS[a,-Abs[b]II, c])/2/.II->I/.HS[a__]:>S[a]//Expand
\end{verbatim}

The analytic continuation of the anomalous dimension for twist-1 operator in ABJM model\footnote{In $\mathcal{N}=4$ SYM theory, we perform analytic continuation for the anomalous dimension of twist-2 operators from $\M_0=2$ to $\M=-1$ to obtain results related to the BFKL equation. For the anomalous dimension of twist-1 operators in ABJM model we have to multiply both values by $2$, because it depends on nested harmonic sums with double argument, $\gamma_{\mathrm{ABJM}}(S)=\sum_{\vec{a}} c_{\vec{a}}H_{\vec{a}}(2S)$.} from $\M_0=2\cdot 2$ to $\M=-2\cdot 1$, which in $\mathcal{N}=4$ SYM theory is related to the BFKL equation, looks like
\begin{eqnarray}
&&\hspace*{-7mm}\gamma_{\mathrm{ABJM}}^{\M=-1,\M_0=4}
=h^2 \left(
-\frac{16}{\omega}
+8\, {\ln\!2}
+\frac{1}{6}\pi ^2 \omega 
- \frac{1}{4}\omega^2 {\z3}
+\frac{1}{1440}\pi ^4 \omega^3 
-\frac{1}{64}\omega^4 {\z5} 
\right)\nonumber\\&&
+h^4 \bigg(
-\frac{128\, {\ln\!2}}{\omega^2}
+\frac{128\, {\ln\!2}^2
+24\, \pi ^2}{\omega}
-\frac{64}{3}{\ln\!2}^3
-24\, \pi ^2{\ln\!2} 
-120\, {\z3}\nonumber\\&&\qquad
+\omega \Big(
\frac{20}{3} \pi ^2 {\ln\!2}^2
+\frac{259}{180}\pi ^4 
-48\, \Li_{-3,1}
+156\, {\ln\!2} {\z3}
\Big)\nonumber\\&&\qquad
+\omega^2 \Big(
-\frac{359}{180} \pi ^4 \ln\!2
-\frac{29}{12}\pi ^2 {\z3} 
-\frac{893}{8}{\z5}
+32\, \Li_{-3,1} {\ln\!2}
+16\, \Li_{-3,1,1}
-54\, {\z3} {\ln\!2}^2
\Big)
\bigg)\nonumber\\&&
+h^6 \bigg(
\frac{-\frac{1024}{3}\pi ^2 
-2048\, {\ln\!2}^2}{\omega^3}
+\frac{\frac{8192}{3}{\ln\!2}^3 
+\frac{3328}{3} \pi ^2 {\ln\!2}
+4224\,{\z3}}{\omega^2}\nonumber\\&&\qquad
+\frac{1}{\omega}\Big(
-\frac{5704}{45}\pi ^4 
+256\, \Li_{-3,1}
-1024\, {\ln\!2}^4
-1152\, \pi ^2{\ln\!2}^2 
-9984\, {\z3}{\ln\!2}\Big)\nonumber\\&&\qquad
+\frac{512}{5}{\ln\!2}^5 
+\frac{4096}{9}\pi ^2 {\ln\!2}^3 
+\frac{2128}{9}\pi ^4{\ln\!2} 
+\frac{1598}{3}\pi ^2 {\z3} 
+\frac{18295}{2}{\z5} \nonumber\\&&\qquad
-1664\, \Li_{-3,1}{\ln\!2} 
+9504\,{\z3}{\ln\!2}^2 \bigg)\,,\label{ABJMADBFKL}
\end{eqnarray}
where $h$ is the effective ABJM QSC coupling constant\footnote{Coupling constant $h$ is a nontrivial function of `t Hooft coupling constant $\lambda$, see Refs.~\cite{Gaiotto:2008cg,Grignani:2008is,Cavaglia:2016ide,Gromov:2014eha}.}.
This results is similar to the expansion of the universal anomalous dimension of twist-2 operators in $\mathcal{N}=4$ SYM theory, since it also has $(a/\omega)^\ell$ behavior (or single logarithms $(a\ln\!x)^\ell$ after inverse Mellin transform). In $\mathcal{N}=4$ SYM theory we have for the analytic continuation of the universal anomalous dimension near $\M=-1+\omega$ we have 
(and similar in QCD for the gluon anomalous dimension near $j=1+\omega$)
\begin{equation}
\gamma_{\mathrm{uni}}^{\M=-1+\omega}=
2\left(\frac{-4g^2}{\omega}\right)
-0\left(\frac{-4g^2}{\omega}\right)^2
+0\left(\frac{-4g^2}{\omega}\right)^3
-4\z3\left(\frac{-4g^2}{\omega}\right)^4+\cdots.\label{N4ADBFKL}
\end{equation}
This result is reproduced from the eigenvalue of the BFKL kernel in the leading-logarithm approximation
\begin{equation}
\frac{\omega}{-4g^2}=\Psi\left(-\frac{\gamma}{2}\right)
+\Psi\left(1+\frac{\gamma}{2}\right)-2\Psi(1)\label{N4BFKL}
\end{equation}
after resolving in $\gamma=\gamma(\omega)$ the following expansion of the above equation
\begin{equation}
\frac{\omega}{-4g^2}=\frac{2}{\gamma}-2\sum_{k=1}^{\infty}\left(\frac{\gamma}{2}\right)^k\zeta(2k+1)\label{N4BFKLExp}
\end{equation}
The same expression (\ref{N4BFKLExp}) can be obtained from the analytic continuation of the simplest harmonic sum near $\M=-1-\omega/2$ plus near $\M=0+\omega/2$, because the analytic continuation of $S_1(N)$ is equal to $\Psi(N+1)-\Psi(1)$ as in Eq.~(\ref{S1}).

The leading poles in the analytically continued anomalous dimension of twist-1 operators in ABJM model~(\ref{ABJMADBFKL}) look like in the case of $\mathcal{N}=4$ SYM theory~(\ref{N4ADBFKL}), but with other coefficients
\begin{equation}
\gamma_{\mathrm{ABJM}}^{\M=-1+\omega}=
2\left(\frac{-8h^2}{\omega}\right)
-2\,\ln\!2\,\left(\frac{-8h^2}{\omega}\right)^2
+4\left(\ln\!2^2+\frac{\pi^2}{6}\right)\left(\frac{-8h^2}{\omega}\right)^3
+\cdots.\label{ABJMADBFKLLLA}
\end{equation}
The term proportional to $\ln\!2$ is contained in the analytic continuation of the sign-alternating analog of $S_1$, the harmonic sum $S_{-1}$. Making the same combination as in Eq.~(\ref{N4BFKL}), we get
\begin{equation}
\frac{\omega}{-8h^2}=
S_{-1}^{AC}(-1-\gamma/2)
+S_{-1}^{AC}(0+\gamma/2)=
\frac{2}{\gamma}
+2\ln\!2
+\frac{\pi^2}{12}\gamma
+\frac{7\pi^4}{2880}\gamma^3+\cdots
\end{equation}
and resolving it with respect to $\gamma$, we obtain
\begin{equation}
\gamma=
2\left(\frac{-8h^2}{\omega}\right)
+4\,\ln\!2\,\left(\frac{-8h^2}{\omega}\right)^2
+\left(4\,\ln\!2^2+\frac{\pi^2}{6}\right)\left(\frac{-8h^2}{\omega}\right)^3
+\cdots\label{ABJMBFKLLLA}
\end{equation}
with the same structure as in Eq.~(\ref{ABJMADBFKLLLA}), but with different coefficients and even with the opposite sign for the second term.
From another side, if we resolve Eq.~(\ref{ABJMADBFKLLLA}) with respect to $\omega$, we get
\begin{equation}
\frac{\omega}{-8h^2}=
\frac{2}{\gamma}
-\ln\!2
+\left(\frac{\ln\!2^2}{2}+\frac{\pi^2}{12}\right)\gamma
+\cdots\,,
\end{equation}
which is the $\gamma$ expansion for the analogue of the BFKL Pomeron eigenvalue. The high-energy scattering in $2+1$-dimensional QCD and the BFKL Pomeron in such model are considered in Refs.~\cite{Li:1994et,Li:1994kx,Ivanov:1998we,Bartels:2004jn}, but we did not find any results relevant to the above expressions.

For the double-logarithmic limit of $\mathcal{N}=4$ SYM theory, which corresponds to the analytical continuation of the anomalous dimension in ABJM model from $\M_0=2\cdot 2$ to $\M=-2\cdot 2$, we have not obtained any pole terms in the $\omega$-expansion, which means that there are no the double logarithms in ABJM model for twist-1 operators.

Another result that can be obtained by means of the analytic continuation is the $\omega$-expansion of anomalous dimension near positive integers values of its argument. Such expansion corresponds to the so-called slope function, first considered in Refs.~\cite{Kotikov:2003fb,Kotikov:2004er} and computed analytically in Ref.~\cite{Basso:2011rs} in $\mathcal{N}=4$ SYM theory. For the three-loop anomalous dimension of twist-1 operators in ABJM model~\cite{Lee:2017mhh,Lee:2019oml}, we found using analytic continuation from $\M_0=0$ to $\M=0$
\begin{eqnarray}
\gamma_{\mathrm{ABJM}}^{\M=0}&\!=&\!
\omega \left(
\frac{\pi ^2}{2} h^2
-\frac{\pi ^4}{4} h^4
+\frac{161\pi^6}{720} h^6 
\right)
+\omega^2 \left(
-\frac{7{\z3}}{4} h^2 
+h^4 \Bigg(
\frac{7 \pi ^2 {\z3}}{12}
-\frac{\pi ^4{\ln\!2} }{4}
+\frac{155 {\z5}}{8}
\right)\nonumber\\&&
+h^6 \left(
\frac{\pi ^6{\ln\!2} }{48}
+\frac{277 \pi ^4 {\z3}}{160}
-\frac{341 \pi ^2 {\z5}}{24}
-\frac{13335 {\z7}}{64}
\right)
\Bigg)\,.
\end{eqnarray}
The firs term of the above $\omega$-expansion is in full agreement with the result from Ref.~\cite{Gromov:2014eha}. The second terms are predictions for the expansion of the next-to-slope function. In principle, it is possible to obtain such expansion for more terms in $\omega$ and for other values of~$\M$.

\section{Conclusion}

In this paper, we presented the algebraic method for the exact expansion of analytically continued nested harmonic sums with real and purely imaginary indices near integer values of their argument. Such nested harmonic sums with purely imaginary indices appeared earlier when calculating the anomalous dimension of the twist-1 operators in ABJM model~\cite{Lee:2017mhh,Lee:2019oml} and BFKL Pomeron eigenvalue in forth order~\cite{Velizhanin:2021bdh}. In the last case, knowing the analytic continuation is the key point in the reconstruction of the general form of BFKL Pomeron eigenvalues, since QCS method~\cite{Gromov:2013pga,Gromov:2014caa,Gromov:2015vua} allows one to compute the pole expansion of the desired result, which have the nested harmonic sums in negative integers. However, the use of the usual harmonic sums with real indices caused a lot of difficulties and the resulting inconsistency, while considering, in addition to the usual sums, nested harmonic sums with purely imaginary indices, allowed to obtain the final result~\cite{Velizhanin:2021bdh}. In the case of ABJM model, we can obtain information about the analytic properties of the anomalous dimension of twist-1 operators known up to the third order~\cite{Lee:2017mhh,Lee:2019oml}. We found that the expansion near negative integer related with the BFKL equation in $\mathcal{N}=4$ SYM theory has the same single-logarithm form as in $\mathcal{N}=4$ SYM theory and in QCD, but with substantially different coefficients.
Moreover, using the proposed method, we computed the $\omega$-expansion of the anomalous dimension of twist-1 operator in ABJM model near $\M=0+\omega$ and found the full agreement with the perturbative expansion of the slope function from Ref.~\cite{Gromov:2014eha}.

We provide the \texttt{MATHEMATICA} code\footnote{This code is available as ancillary files of \texttt{arXiv}-version of this paper and on GitHub \url{https://github.com/vitvel/ACHSI}.} that perform analytic continuation of nested harmonic sums $S_{\vec{a}}$ or $\Psi_{\vec{a}}$ function with real and purely imaginary indices near negative and positive integers. The proposed method should also work for other generalisation of usual harmonic sums, such as the so-called \texttt{SSum}~\cite{Moch:2001zr} or cyclotomic sums~\cite{Ablinger:2011te}, but this would require a database for the relations between the corresponding generalization of the multiple zeta values or multiple polylogarithms.


 \subsection*{Acknowledgments}

The computations of the database for the relations between multiple polylogarithms were performed with resources provided by the PIK Data Centre in PNPI NRC "Kurchatov Institute".

I would like to thank M. Kalmykov and A. Shuvaev for useful discussions.

The research was supported by a grant from the Russian Science Foundation No. 22-22-00803, https://rscf.ru/en/project/22-22-00803/.


\begin{thebibliography}{10}

\bibitem{Vermaseren:1998uu}
J.~A.~M. Vermaseren, {\it {Harmonic sums, Mellin transforms and integrals}},
  {\em Int. J. Mod. Phys. A} {\bf 14} (1999) 2037--2076,
  [\href{http://xxx.lanl.gov/abs/hep-ph/9806280}{{\tt hep-ph/9806280}}].

\bibitem{GonzalezArroyo:1979df}
A.~Gonzalez-Arroyo, C.~Lopez, and F.~J. Yndurain, {\it {Second Order
  Contributions to the Structure Functions in Deep Inelastic Scattering. 1.
  Theoretical Calculations}},  {\em Nucl. Phys. B} {\bf 153} (1979) 161--186.

\bibitem{Kotikov:2005gr}
A.~V. Kotikov and V.~N. Velizhanin, {\it {Analytic continuation of the Mellin
  moments of deep inelastic structure functions}},
  \href{http://xxx.lanl.gov/abs/hep-ph/0501274}{{\tt hep-ph/0501274}}.

\bibitem{Lipatov:1976zz}
L.~N. Lipatov, {\it {Reggeization of the Vector Meson and the Vacuum
  Singularity in Nonabelian Gauge Theories}},  {\em Sov. J. Nucl. Phys.} {\bf
  23} (1976) 338--345.

\bibitem{Kuraev:1977fs}
E.~A. Kuraev, L.~N. Lipatov, and V.~S. Fadin, {\it {The Pomeranchuk Singularity
  in Nonabelian Gauge Theories}},  {\em Sov. Phys. JETP} {\bf 45} (1977)
  199--204.

\bibitem{Balitsky:1978ic}
I.~I. Balitsky and L.~N. Lipatov, {\it {The Pomeranchuk Singularity in Quantum
  Chromodynamics}},  {\em Sov. J. Nucl. Phys.} {\bf 28} (1978) 822--829.

\bibitem{Gorshkov:1966qd}
V.~G. Gorshkov, V.~N. Gribov, L.~N. Lipatov, and G.~V. Frolov, {\it {Double
  logarithmic asymptotics of quantum electrodynamics}},  {\em Phys. Lett.} {\bf
  22} (1966) 671--673.

\bibitem{Kirschner:1983di}
R.~Kirschner and L.~n. Lipatov, {\it {Double Logarithmic Asymptotics and Regge
  Singularities of Quark Amplitudes with Flavor Exchange}},  {\em Nucl. Phys.
  B} {\bf 213} (1983) 122--148.

\bibitem{Gromov:2015vua}
N.~Gromov, F.~Levkovich-Maslyuk, and G.~Sizov, {\it {Pomeron Eigenvalue at
  Three Loops in $\mathcal N=$ 4 Supersymmetric Yang-Mills Theory}},  {\em
  Phys. Rev. Lett.} {\bf 115} (2015), no.~25 251601,
  [\href{http://xxx.lanl.gov/abs/1507.0401}{{\tt arXiv:1507.0401}}].

\bibitem{Velizhanin:2015xsa}
V.~N. Velizhanin, {\it {BFKL pomeron in the next-to-next-to-leading
  approximation in the planar N=4 SYM theory}},
  \href{http://xxx.lanl.gov/abs/1508.0285}{{\tt arXiv:1508.0285}}.

\bibitem{Velizhanin:2021bdh}
V.~N. Velizhanin, {\it {NNNLLA BFKL pomeron eigenvalue in the planar N=4 SYM
  theory}},  \href{http://xxx.lanl.gov/abs/2106.0652}{{\tt arXiv:2106.0652}}.

\bibitem{Velizhanin:2011pb}
V.~N. Velizhanin, {\it {Double-logs, Gribov-Lipatov reciprocity and wrapping}},
   {\em JHEP} {\bf 08} (2011) 092,
  [\href{http://xxx.lanl.gov/abs/1104.4100}{{\tt arXiv:1104.4100}}].

\bibitem{Velizhanin:2014dia}
V.~N. Velizhanin, {\it {Generalised double-logarithmic equation in QCD}},  {\em
  Mod. Phys. Lett. A} {\bf 32} (2017), no.~38 1750213,
  [\href{http://xxx.lanl.gov/abs/1412.7143}{{\tt arXiv:1412.7143}}].

\bibitem{Velizhanin:2022seo}
V.~N. Velizhanin, {\it {Exact result in $ \mathcal{N} $ = 4 SYM theory:
  generalised double-logarithmic equation}},  {\em JHEP} {\bf 05} (2022) 176,
  [\href{http://xxx.lanl.gov/abs/2201.0461}{{\tt arXiv:2201.0461}}].

\bibitem{Albino:2009ci}
S.~Albino, {\it {Analytic Continuation of Harmonic Sums}},  {\em Phys. Lett. B}
  {\bf 674} (2009) 41--48, [\href{http://xxx.lanl.gov/abs/0902.2148}{{\tt
  arXiv:0902.2148}}].

\bibitem{Blumlein:2009ta}
J.~Blumlein, {\it {Structural Relations of Harmonic Sums and Mellin Transforms
  up to Weight w = 5}},  {\em Comput. Phys. Commun.} {\bf 180} (2009)
  2218--2249, [\href{http://xxx.lanl.gov/abs/0901.3106}{{\tt
  arXiv:0901.3106}}].

\bibitem{Velizhanin:2020avm}
V.~N. Velizhanin, {\it {Analytic continuation of harmonic sums near the integer
  values}},  {\em Int. J. Mod. Phys. A} {\bf 35} (2020), no.~33 2050210.

\bibitem{Remiddi:1999ew}
E.~Remiddi and J.~A.~M. Vermaseren, {\it {Harmonic polylogarithms}},  {\em Int.
  J. Mod. Phys. A} {\bf 15} (2000) 725--754,
  [\href{http://xxx.lanl.gov/abs/hep-ph/9905237}{{\tt hep-ph/9905237}}].

\bibitem{Vermaseren:2000nd}
J.~A.~M. Vermaseren, {\it {New features of FORM}},
  \href{http://xxx.lanl.gov/abs/math-ph/0010025}{{\tt math-ph/0010025}}.

\bibitem{Kuipers:2012rf}
J.~Kuipers, T.~Ueda, J.~A.~M. Vermaseren, and J.~Vollinga, {\it {FORM version
  4.0}},  {\em Comput. Phys. Commun.} {\bf 184} (2013) 1453--1467,
  [\href{http://xxx.lanl.gov/abs/1203.6543}{{\tt arXiv:1203.6543}}].

\bibitem{Ruijl:2017dtg}
B.~Ruijl, T.~Ueda, and J.~Vermaseren, {\it {FORM version 4.2}},
  \href{http://xxx.lanl.gov/abs/1707.0645}{{\tt arXiv:1707.0645}}.

\bibitem{Blumlein:2009cf}
J.~Blumlein, D.~J. Broadhurst, and J.~A.~M. Vermaseren, {\it {The Multiple Zeta
  Value Data Mine}},  {\em Comput. Phys. Commun.} {\bf 181} (2010) 582--625,
  [\href{http://xxx.lanl.gov/abs/0907.2557}{{\tt arXiv:0907.2557}}].

\bibitem{Velizhanin:2022ays}
V.~N. Velizhanin, {\it {Analytic continuation of harmonic sums: Dispersion
  representation}},  {\em Nucl. Phys. B} {\bf 984} (2022) 115976,
  [\href{http://xxx.lanl.gov/abs/2205.1518}{{\tt arXiv:2205.1518}}].

\bibitem{Lee:2017mhh}
R.~N. Lee and A.~I. Onishchenko, {\it {ABJM quantum spectral curve and Mellin
  transform}},  {\em JHEP} {\bf 05} (2018) 179,
  [\href{http://xxx.lanl.gov/abs/1712.0041}{{\tt arXiv:1712.0041}}].

\bibitem{Lee:2019oml}
R.~N. Lee and A.~I. Onishchenka, {\it {ABJM quantum spectral curve at twist 1:
  algorithmic perturbative solution}},  {\em JHEP} {\bf 11} (2019) 018,
  [\href{http://xxx.lanl.gov/abs/1905.0311}{{\tt arXiv:1905.0311}}].

\bibitem{Moch:2001zr}
S.~Moch, P.~Uwer, and S.~Weinzierl, {\it {Nested sums, expansion of
  transcendental functions and multiscale multiloop integrals}},  {\em J. Math.
  Phys.} {\bf 43} (2002) 3363--3386,
  [\href{http://xxx.lanl.gov/abs/hep-ph/0110083}{{\tt hep-ph/0110083}}].

\bibitem{Ablinger:2011te}
J.~Ablinger, J.~Blumlein, and C.~Schneider, {\it {Harmonic Sums and
  Polylogarithms Generated by Cyclotomic Polynomials}},  {\em J. Math. Phys.}
  {\bf 52} (2011) 102301, [\href{http://xxx.lanl.gov/abs/1105.6063}{{\tt
  arXiv:1105.6063}}].

\bibitem{Goncharov:MPL}
A.~B. Goncharov, {\it {Multiple polylogarithms, cyclotomy and modular
  complexes}},  {\em Mathematical Research Letters} {\bf 5} (1998) 497--516,
  [\href{http://xxx.lanl.gov/abs/1105.2076}{{\tt arXiv:1105.2076}}].

\bibitem{Velizhanin:I}
V.~Velizhanin, {\it {BFKL amplitudes and multiple polylogarithm values at
  fourth roots of unity}}, in preparation.

\bibitem{Bauer:2000cp}
C.~W. Bauer, A.~Frink, and R.~Kreckel, {\it {Introduction to the GiNaC
  framework for symbolic computation within the C++ programming language}},
  {\em J. Symb. Comput.} {\bf 33} (2002) 1--12,
  [\href{http://xxx.lanl.gov/abs/cs/0004015}{{\tt cs/0004015}}].

\bibitem{Vollinga:2004sn}
J.~Vollinga and S.~Weinzierl, {\it {Numerical evaluation of multiple
  polylogarithms}},  {\em Comput. Phys. Commun.} {\bf 167} (2005) 177,
  [\href{http://xxx.lanl.gov/abs/hep-ph/0410259}{{\tt hep-ph/0410259}}].

\bibitem{Ferguson1999AnalysisOP}
H.~R.~P. Ferguson, D.~H. Bailey, and S.~Arno, {\it {Analysis of PSLQ, an
  integer relation finding algorithm}},  {\em Math. Comput.} {\bf 68} (1999)
  351--369.

\bibitem{Zimmermann:CPSLQ}
P.~Zimmermann, {\it {An implementation of PSLQ in GMP}}, .
  https://members.loria.fr/PZimmermann/free/.

\bibitem{Gromov:2013pga}
N.~Gromov, V.~Kazakov, S.~Leurent, and D.~Volin, {\it {Quantum Spectral Curve
  for Planar $\mathcal{N} = 4$ Super-Yang-Mills Theory}},  {\em Phys. Rev.
  Lett.} {\bf 112} (2014), no.~1 011602,
  [\href{http://xxx.lanl.gov/abs/1305.1939}{{\tt arXiv:1305.1939}}].

\bibitem{Gromov:2014caa}
N.~Gromov, V.~Kazakov, S.~Leurent, and D.~Volin, {\it {Quantum spectral curve
  for arbitrary state/operator in AdS$_{5}$/CFT$_{4}$}},  {\em JHEP} {\bf 09}
  (2015) 187, [\href{http://xxx.lanl.gov/abs/1405.4857}{{\tt
  arXiv:1405.4857}}].

\bibitem{Aharony:2008ug}
O.~Aharony, O.~Bergman, D.~L. Jafferis, and J.~Maldacena, {\it {N=6
  superconformal Chern-Simons-matter theories, M2-branes and their gravity
  duals}},  {\em JHEP} {\bf 10} (2008) 091,
  [\href{http://xxx.lanl.gov/abs/0806.1218}{{\tt arXiv:0806.1218}}].

\bibitem{Gaiotto:2008cg}
D.~Gaiotto, S.~Giombi, and X.~Yin, {\it {Spin Chains in N=6 Superconformal
  Chern-Simons-Matter Theory}},  {\em JHEP} {\bf 04} (2009) 066,
  [\href{http://xxx.lanl.gov/abs/0806.4589}{{\tt arXiv:0806.4589}}].

\bibitem{Grignani:2008is}
G.~Grignani, T.~Harmark, and M.~Orselli, {\it {The SU(2) x SU(2) sector in the
  string dual of N=6 superconformal Chern-Simons theory}},  {\em Nucl. Phys. B}
  {\bf 810} (2009) 115--134, [\href{http://xxx.lanl.gov/abs/0806.4959}{{\tt
  arXiv:0806.4959}}].

\bibitem{Cavaglia:2016ide}
A.~Cavagli\`a, N.~Gromov, and F.~Levkovich-Maslyuk, {\it {On the Exact
  Interpolating Function in ABJ Theory}},  {\em JHEP} {\bf 12} (2016) 086,
  [\href{http://xxx.lanl.gov/abs/1605.0488}{{\tt arXiv:1605.0488}}].

\bibitem{Gromov:2014eha}
N.~Gromov and G.~Sizov, {\it {Exact Slope and Interpolating Functions in N=6
  Supersymmetric Chern-Simons Theory}},  {\em Phys. Rev. Lett.} {\bf 113}
  (2014), no.~12 121601, [\href{http://xxx.lanl.gov/abs/1403.1894}{{\tt
  arXiv:1403.1894}}].

\bibitem{Li:1994et}
M.~Li and C.-I. Tan, {\it {High-energy scattering in (2+1) QCD}},  {\em Phys.
  Rev. D} {\bf 50} (1994) 1140--1149,
  [\href{http://xxx.lanl.gov/abs/hep-th/9401134}{{\tt hep-th/9401134}}].

\bibitem{Li:1994kx}
M.~Li and C.-I. Tan, {\it {High-energy scattering in (2+1) QCD: A Dipole
  picture}},  {\em Phys. Rev. D} {\bf 51} (1995) 3287--3297,
  [\href{http://xxx.lanl.gov/abs/hep-ph/9407299}{{\tt hep-ph/9407299}}].

\bibitem{Ivanov:1998we}
D.~Y. Ivanov, R.~Kirschner, E.~M. Levin, L.~N. Lipatov, L.~Szymanowski, and
  M.~Wusthoff, {\it {The BFKL pomeron in (2+1)-dimensional QCD}},  {\em Phys.
  Rev. D} {\bf 58} (1998) 074010,
  [\href{http://xxx.lanl.gov/abs/hep-ph/9804443}{{\tt hep-ph/9804443}}].

\bibitem{Bartels:2004jn}
J.~Bartels, V.~S. Fadin, and L.~N. Lipatov, {\it {Solution of the fan diagram
  equation in (2+1)-dimensional QCD}},  {\em Nucl. Phys. B} {\bf 698} (2004)
  255--276, [\href{http://xxx.lanl.gov/abs/hep-ph/0406193}{{\tt
  hep-ph/0406193}}].

\bibitem{Kotikov:2003fb}
A.~V. Kotikov, L.~N. Lipatov, and V.~N. Velizhanin, {\it {Anomalous dimensions
  of Wilson operators in N=4 SYM theory}},  {\em Phys. Lett. B} {\bf 557}
  (2003) 114--120, [\href{http://xxx.lanl.gov/abs/hep-ph/0301021}{{\tt
  hep-ph/0301021}}].

\bibitem{Kotikov:2004er}
A.~V. Kotikov, L.~N. Lipatov, A.~I. Onishchenko, and V.~N. Velizhanin, {\it
  {Three loop universal anomalous dimension of the Wilson operators in $N=4$
  SUSY Yang-Mills model}},  {\em Phys. Lett. B} {\bf 595} (2004) 521--529,
  [\href{http://xxx.lanl.gov/abs/hep-th/0404092}{{\tt hep-th/0404092}}].
  [Erratum: Phys.Lett.B 632, 754--756 (2006)].

\bibitem{Basso:2011rs}
B.~Basso, {\it {An exact slope for AdS/CFT}},
  \href{http://xxx.lanl.gov/abs/1109.3154}{{\tt arXiv:1109.3154}}.

\end{thebibliography}
\end{document}